%% file: main.tex
\PassOptionsToPackage{prologue,table,xcdraw}{xcolor}
\documentclass[conference]{IEEEtran}
\IEEEoverridecommandlockouts

\usepackage{etoolbox}
\makeatletter
\patchcmd{\@makecaption}
  {\scshape}
  {}
  {}
  {}
\makeatletter
\patchcmd{\@makecaption}
  {\\}
  {.\ }
  {}
  {}
\makeatother
\def\tablename{Table}
\usepackage{caption}

\usepackage{cite}
\usepackage{amsmath,amssymb,amsfonts}
\usepackage{graphicx}
\usepackage{textcomp}
\usepackage{balance}
\usepackage{bm}
\usepackage{url}
\usepackage[table,xcdraw]{xcolor}
\usepackage{algorithm}
\usepackage{algpseudocode}

\usepackage{enumitem}
\usepackage{multirow}
\usepackage[normalem]{ulem}
\useunder{\uline}{\ul}{}

\usepackage{graphicx}

\newcommand{\ie}{\emph{i.e., }}
\newcommand{\eg}{\emph{e.g., }}

\newcommand{\aka}

\def\BibTeX{{\rm B\kern-.05em{\sc i\kern-.025em b}\kern-.08em
    T\kern-.1667em\lower.7ex\hbox{E}\kern-.125emX}}
\begin{document}

\title{DiscRec: Disentangled Semantic–Collaborative Modeling for Generative Recommendation}

\author{\IEEEauthorblockN{Chang Liu\IEEEauthorrefmark{1}, Yimeng Bai\IEEEauthorrefmark{2}, Xiaoyan Zhao\IEEEauthorrefmark{3}, Yang Zhang\IEEEauthorrefmark{4}, Fuli Feng\IEEEauthorrefmark{2}, and Wenge Rong\IEEEauthorrefmark{1}}
\IEEEauthorblockA{
\IEEEauthorrefmark{1}Beihang University, Beijing, China\\
\IEEEauthorrefmark{2}University of Science and Technology of China, Hefei, China\\
\IEEEauthorrefmark{3}The Chinese University of Hong Kong, Hong Kong, China\\
\IEEEauthorrefmark{4}National University of Singapore, Singapore, Singapore\\
chang.liu@buaa.edu.cn, baiyimeng@mail.ustc.edu.cn, xzhao@se.cuhk.edu.hk,\\
zyang1580@gmail.com, fulifeng93@gmail.com, w.rong@buaa.edu.cn}
\vspace{-1em}
}

\maketitle

\begin{abstract}
\input{1_abs}
\end{abstract}


\section{Introduction}
\input{2_intro}

\section{Preliminary}
\input{3_pre}

\section{Methodology}
\input{4_method}

\section{Experiment}
\input{5_exp}

\vspace{-5pt}
\section{Related Work}
\input{6_rel}

\section{Conclusion}
\input{7_con}

\section{AI-Generated Content Acknowledgement}
\input{8_ai}

\bibliographystyle{IEEEtran}
\balance
\bibliography{9_ref}

\end{document}

%% file: 1_abs.tex
\label{abs}
Generative recommendation is emerging as a powerful paradigm that directly generates item predictions, moving beyond traditional matching-based approaches. However, current methods face two key challenges: token-item misalignment, where uniform token-level modeling ignores item-level granularity that is critical for collaborative signal learning, and semantic–collaborative signal entanglement, where collaborative and semantic signals exhibit distinct distributions yet are fused in a unified embedding space, leading to conflicting optimization objectives that limit the recommendation performance.

To address these issues, we propose DiscRec, a novel framework that enables \underline{Di}sentangled \underline{S}emantic-\underline{C}ollaborative signal modeling with flexible fusion for generative \underline{Rec}ommendation. 
First, DiscRec introduces item-level position embeddings, assigned based on indices within each semantic ID, enabling explicit modeling of item structure in input token sequences. 
Second, DiscRec employs a dual-branch module to disentangle the two signals at the embedding layer: a semantic branch encodes semantic signals using original token embeddings, while a collaborative branch applies localized attention restricted to tokens within the same item to effectively capture collaborative signals. A gating mechanism subsequently fuses both branches while preserving the model’s ability to model sequential dependencies. Extensive experiments on four real-world datasets demonstrate that DiscRec effectively decouples these signals and consistently outperforms state-of-the-art baselines. Our codes are available on \url{https://github.com/Ten-Mao/DiscRec}.

%% file: 2_intro.tex
\label{intro}

Generative recommendation, emerging as a promising next-generation paradigm, has attracted increasing attention from both industry and academia~\cite{GeneRec,OneRec,HSTU,PinRec,MTGR,UniGRF}, as it departs from the traditional discriminative query-candidate matching framework by directly generating the next item in an autoregressive manner. Typically, this paradigm involves two stages: (1) a \textit{tokenization stage}, where a tokenizer model employs discretization methods (\eg RQ-VAE~\cite{RQ-VAE}) to convert the specific representation of items into corresponding semantic IDs~\cite{TIGER,LETTER,ETEGRec,SEATER,TokenRec}; (2) a \textit{recommendation stage}, where a Transformer-based recommender model (\eg T5~\cite{T5}) autoregressively generates item predictions based on the semantic ID sequences constructed from user interaction histories.

\begin{figure}[t]
    \centering
    \includegraphics[width=\linewidth]{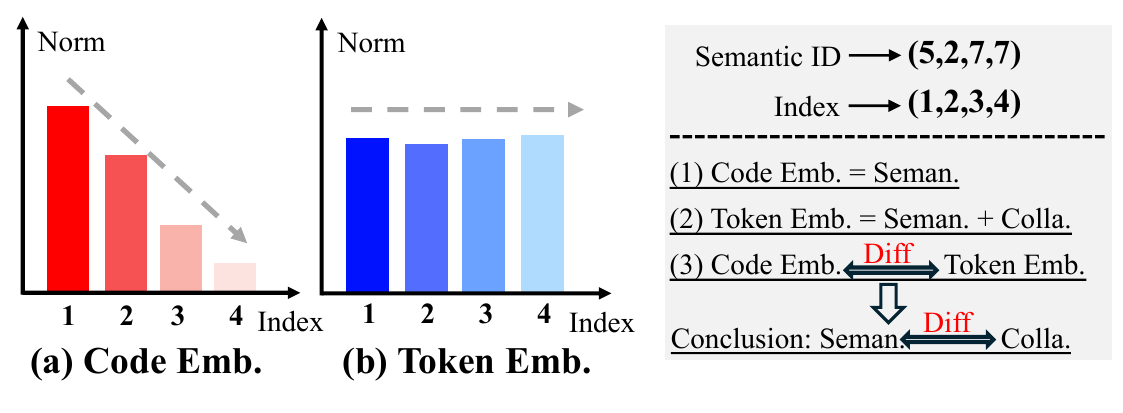}
    \caption{Comparison of the distribution between code embeddings (\ie entries from the RQ-VAE~\cite{RQ-VAE} tokenizer's codebook) and token embeddings (\ie entries from the T5~\cite{T5} recommender embedding table) across indices of the semantic ID, using embedding norm as a measure of information capacity and averaged over all items.(a) Code embeddings show a clear hierarchical decay, while (b) token embeddings have a more uniform distribution. The right side illustrates how code embeddings primarily encode semantic signals, while token embeddings incorporate both semantic and collaborative signals. This difference reflects the distinct distributions of these two signal types.}
    \label{fig:code_token}
\end{figure}

Within this paradigm, the recommender undertakes two core tasks: extracting \textit{semantic signals} embedded in the semantic IDs, and capturing \textit{collaborative signals} through the modeling of sequential transition patterns in user interaction histories. Therefore, superior generative recommendations fundamentally relies on the seamless integration of these two heterogeneous information sources. From this perspective, we identify two key challenges: (1) 
\textbf{Token-item misalignment.} Existing generative recommender models \cite{TIGER,LETTER,EAGER} treat all tokens uniformly, ignoring the item boundaries they belong to. This leads to the token-item misalignment, where the lack of item-level granularity undermines the model’s ability to effectively learn collaborative signals.
(2) \textbf{Semantic–collaborative signal entanglement.} These models \cite{TIGER,LETTER} often entangle collaborative and semantic signals within a unified token embedding space. However, as shown in Figure~\ref{fig:code_token}, our empirical analysis reveals that these two signals follow distinct distribution patterns across indices of the semantic ID—semantic signals exhibit a hierarchical decay, whereas collaborative signals are not. This discrepancy introduces imcompatible optimization objectives, which not only leads to representational interference during training, but also degrades the model's ability to effectively capture either signal—ultimately limiting recommendation performance.

Existing works on generative recommendation primarily focus on leveraging both collaborative and semantic signals but do not directly address the underlying challenges. The first direction focuses on the tokenization stage, where semantic IDs are enhanced through the incorporation of auxiliary collaborative regularization losses~\cite{LETTER,ETEGRec,ColaRec,PRORec}. However, it relies on pretrained collaborative embeddings and remains constrained by the downstream recommender’s misaligned token-level architecture. 
The second direction operates at the recommendation stage, introducing a two-stream architecture that builds entirely separate models for semantic and collaborative signals~\cite{EAGER,SC-Rec,EAGER-LLM,TSSR}. While this achieves complete disentanglement at the embedding level, it requires maintaining two isolated token spaces and model flows, which substantially increases computational cost. Moreover, the two streams interact only at the final reranking stage, preventing deep integration and limiting the exploration of their complementary strengths.

Building upon the limitations of prior approaches \cite{EAGER, LETTER}, we distill our target for generative recommender models as achieving \textbf{item-aware alignment} and \textbf{semantic-collaborative signal disentanglement with flexible fusion}. Initially, we aim to incorporate item awareness to align the modeling granularity with the item-level nature of recommendation tasks. Based on this foundation, we further introduce a flexible disentanglement mechanism that separates semantic and collaborative signals explicitly at the token embedding layer to mitigate representational interference during optimization. In contrast to two-stream architectures enforcing strict separation throughout the model, we retain cross-signal interactions at deeper layers, allowing the model to fully leverage their complementary strengths for improved recommendation performance.

To this end, we propose DiscRec, a novel framework that enables \underline{Di}sentangled \underline{S}emantic-\underline{C}ollaborative signal modeling with flexible fusion for generative \underline{Rec}ommendation. First, to achieve item-aware alignment, we introduce item-level position embeddings for each token, with positions assigned according to their indices within the corresponding semantic ID sequence.
By sharing these embeddings across different items, we enable the model to efficiently discern the item-level structure within the input token sequences. 
Second, to enable semantic-collaborative signal disentanglement with flexible fusion, we propose a dual-branch module within a single workflow. The semantic branch directly models semantic signals using the original token embeddings. In parallel, the collaborative branch combines token embeddings with item-level position embeddings and employs a Transformer with localized attention, where attention is restricted to tokens within the same item, to effectively capture collaborative signals. The outputs from both branches are adaptively fused via a gating mechanism, allowing the disentanglement to be seamlessly integrated in a plug-and-play manner at the embedding layer, while preserving the original architecture’s ability to model complex sequential dependencies.
To demonstrate the generalizability of  DiscRec, we incorporate it into two representative generative recommendation frameworks, TIGER~\cite{TIGER} and LETTER~\cite{LETTER}, and systematically validate it through experiments on four real-world datasets. Experimental results demonstrate that DiscRec successfully decouples collaborative and semantic signals and consistently achieves superior recommendation performance.

The main contributions are summarized as follows:
\begin{itemize}[leftmargin=*]
\item We empirically demonstrate that collaborative and semantic signals exhibit significantly different distributions, and we identify two critical challenges in generative recommendation: 
token-item misalignment, semantic–collaborative signal entanglement.
\item We propose DiscRec, a novel framework that integrates item-level position embeddings with a dual-branch module to achieve item-aware alignment and semantic-collaborative
signal disentanglement with flexible fusion.
\item We conduct extensive experiments on multiple real-world datasets, validating the effectiveness of DiscRec in enhancing generative recommendation performance.
\end{itemize}

%% file: 3_pre.tex
\label{pre}

\begin{figure}[t]
    \centering
    \includegraphics[width=0.45\textwidth]{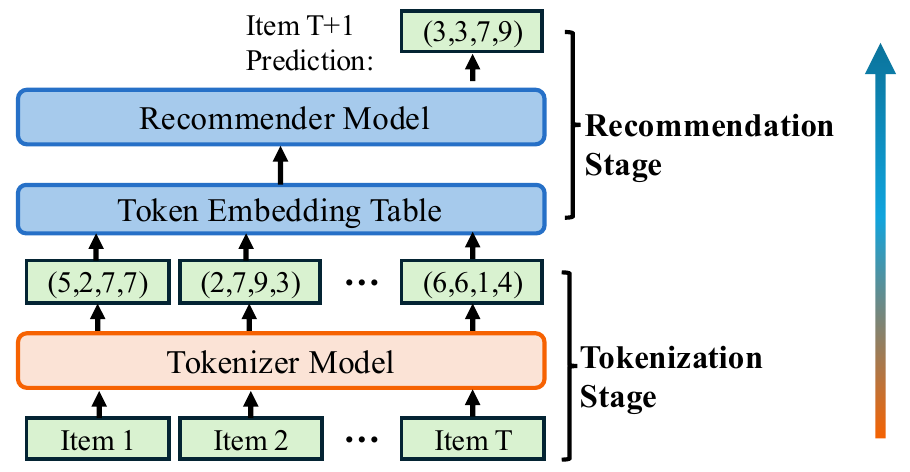}
    \caption{Overview of the generative recommendation paradigm, which sequentially applies the tokenization stage to discrete the semantic embeddings, and the recommendation stage to generate item predictions. Each tuple (\eg (5,2,7,7)) denotes a sequence of tokens representing a semantic ID of an item.}
    \label{fig:GR}
\end{figure}

In this section, we formally define the task of generative recommendation. Consistent with prior studies~\cite{TIGER, LETTER}, we focus on the sequential recommendation setting. Given a universal item set \(\mathcal{I}\) and a user’s historical interaction sequence \(S = [i_1, i_2, \ldots, i_T]\), the objective is to predict the next item \(i_{T+1} \in \mathcal{I}\) that the user is likely to interact with. Departing from the traditional query-candidate matching paradigm~\cite{SASRec}, generative recommendation reframes the task as directly generating the next item. As shown in Figure~\ref{fig:GR}, this paradigm typically consists of two core components:

\subsection{Tokenization Stage}
\subsubsection{Task Formulation}
At this stage, each item is encoded into a sequence of discrete tokens by feeding its semantic embedding into a tokenizer model $\mathcal{T}$. Following TIGER~\cite{TIGER}, a representative generative recommendation framework, we adopt the Residual Vector Quantized Variational Autoencoder (RQ-VAE)~\cite{RQ-VAE} to discretize item representations into semantic IDs. This hierarchical encoding offers two primary advantages. First, it inherently induces a tree-structured item space, which is well-suited for generative modeling. Second, items sharing common prefix tokens naturally capture collaborative semantics, enabling effective information sharing among semantically related items. Generally, given an item $i$, the tokenizer $\mathcal{T}$ takes its semantic embedding $\bm{z}$—typically a pretrained text embedding and produces a sequence of quantized tokens across $L$ hierarchical levels, formulated as:
\begin{equation}
[c_1,\ldots, c_L] = \mathcal{T}(\bm{z}),
\end{equation}
where $c_l$ represents the token assigned to item $i$ at the $l$-th level of the hierarchy. Then, each level uses a separate codebook to incrementally refine the representation by encoding residuals from the previous level.

\vspace{+5pt}
\subsubsection{Model Optimization}
We now describe the process of obtaining hierarchical tokens for a specific item $i$. We begin by encoding the input semantic embedding $\bm{z}$ into a latent representation using a MLP-based encoder:
\begin{equation}
\bm{r} = \text{Encoder}_\mathcal{T}(\bm{z}).
\end{equation}
The latent representation $\bm{r}$ is then quantized into $L$ discrete tokens by querying $L$ distinct codebooks. At each level $l$, we have a codebook $\bm{\mathcal{C}}_{l}=\{\bm{e}_l^k\}_{k=1}^{K}$, where each $e_l^k$ is a learnable embedding vector and $K$ is the codebook size. The codebook acts as a discrete vocabulary that maps continuous latent vectors into symbolic tokens by selecting the nearest code in Euclidean space.
Specifically, this quantization process is performed as follows:
\begin{align}
c_l &= \underset{k}{\arg\min} \; \lVert \bm{v}_l - \bm{e}_l^k \rVert^2, \\
\bm{v}_l &= \bm{v}_{l-1} - \bm{e}_{l-1}^{c_{l-1}}.
\end{align}
Here, $\bm{v}_l$ denotes the residual vector at the $l$-th level ($\bm{v}_1=\bm{r}$), and the assignment is based on the Euclidean distance between $\bm{v}_l$ and each entry in the corresponding codebook. After quantizing the initial semantic embedding into a hierarchy of tokens from coarse to fine granularity, we obtain the quantized representation $\tilde{\bm{r}}$, which is then fed into an MLP-based decoder to reconstruct the semantic embedding:
\begin{align}
\tilde{\bm{r}} &= \sum_{l=1}^{L} \bm{e}_{l}^{c_l},\\
\tilde{\bm{z}} &= \text{Decoder}_{\mathcal{T}}(\tilde{\bm{r}}).
\end{align}
Overall, the tokenization loss for item $i$ is defined as\footnote{Note that additional auxiliary losses, such as collaborative regularization, may also be included (see LETTER~\cite{LETTER}).}:
\begin{equation}\label{eq:loss_token}
\mathcal{L}_\mathcal{T} = \lVert \tilde{\bm{z}} - \bm{z} \rVert^2 + \sum_{l=1}^{L} (\lVert \text{sg}[\bm{v}_l] - \bm{e}_l^{c_l} \rVert^2 + \beta \lVert \bm{v}_l - \text{sg}[\bm{e}_l^{c_l}] \rVert^2).
\end{equation}
The first term represents a reconstruction loss, which encourages the reconstructed embedding \( \tilde{\bm{z}} \) to closely approximate the original input \( \bm{z} \). The remaining two terms correspond to quantization losses, aiming to minimize the discrepancy between the residual vectors and their associated codebook entries. Here, \( \text{sg}[\cdot] \) denotes the stop-gradient operation~\cite{RQ-VAE}, and $\beta$ is a weighting coefficient that balances the learning dynamics between the encoder and the codebooks.

By apply the discretization process for each item in user history, the input interaction sequence \(S\) can be transformed into a token sequence:
\begin{equation}
X = [c_1^1, c_2^1, \ldots, c_L^1, c_{1}^{T}, \ldots, c_{L}^T], 
\end{equation}
where \(c_l^j\) denotes the \(l\)-th token of item \(i_j\) (\(j \in \{1, \ldots, T\}\)), \(T\) is the length of the original interaction sequence \(S\), and \(L\) is the length of each item identifier. The ground-truth next item \(i_{T+1}\) is similarly represented as a token sequence:
\begin{equation}
Y = [c_1^{T+1}, \ldots, c_L^{T+1}].
\end{equation}

\subsection{Recommendation Stage}
\subsubsection{Task Formulation}
At this stage, the next-item prediction task is framed as a sequence generation problem over the tokenized interaction history. Specifically, each token of the target item is generated autoregressively, conditioned on the input token sequence and all previously generated tokens. This approach enables the extraction of semantic signals from the semantic IDs produced during tokenization, while capturing collaborative signals through modeling the sequential transition patterns in user interactions.

\vspace{+5pt}
\subsubsection{Model Optimization}
Consistent with prior work~\cite{TIGER, LETTER}, our recommender model \(\mathcal{R}\) employs an encoder-decoder architecture based on T5~\cite{T5}. For the tokenized sequence $X$, we append a end-of-sequence token [EOS] and then construct its embedding sequence $\bm{E}^X\in \mathbb{R}^{(LT+1)\times D}$ by looking up the token embedding table $\bm{\mathcal{O}}$, where $D$ denotes the embedding dimension. The embedding sequence of $X$ is defined as:
\begin{equation}\label{eq:token_emb_X}
\bm{E}^X = [\bm{o}_1^1, \bm{o}_2^1, \ldots, \bm{o}_L^T, \bm{o}^{\text{EOS}}],
\end{equation}
where $\bm{o}_l^t$ denotes the embedding of the $l$-th token for item $i_t$ (\ie token $c_l^t$). 
This embedding sequence is then fed into the Transformer encoder to obtain the corresponding hidden representation:
\begin{equation}\label{eq:hidden_encoder}
\bm{H}^{\text{Enc}}=\text{Encoder}_{\mathcal{R}}(\bm{E}^X).
\end{equation}
For decoding, a special beginning-of-sequence token [BOS] is prepended to the target token sequence $Y$, resulting in an embedding sequence $\bm{E}^Y\in\mathbb{R}^{(1+L)\times D}$ defined as:
\begin{equation}\label{eq:token_emb_Y}
\bm{E}^Y=[\bm{o}^{\text{BOS}}, \bm{o}_1^{T+1},\ldots,\bm{o}_{L}^{T+1}].
\end{equation}
The Transformer decoder then takes the encoder output $\bm{H}^{\text{Enc}}$ and the target embedding sequence $\bm{E}^Y$ as input to model the user preference representation:
\begin{equation}\label{eq:hidden_decoder}
\bm{H}^{\text{Dec}}=\text{Decoder}_{\mathcal{R}}(\bm{H}^{\text{Enc}},\bm{E}^Y).
\end{equation}
The decoder output $\bm{H}^{\text{Dec}}$ is projected onto the token embedding space via an inner product with the token embedding table $\mathcal{O}$ to predict the target tokens. The recommendation loss is defined as the negative log-likelihood of the target tokens under the sequence-to-sequence learning framework:
\begin{equation}\label{eq:loss_rec}
\mathcal{L}_\mathcal{R}=-\sum_{l=1}^{L}\log P(Y_l|X,Y_{<l}),
\end{equation}
where $Y_l$ denotes the $l$-th token in the target sequence, and $Y_{<l}$ represents all preceding tokens.

\subsection{Distribution Analysis of Semantic and Collaborative Signals}
To further elucidate our core motivation, we offer a more detailed explanation of the experiment shown in Figure~\ref{fig:code_token}, with the aim of demonstrating the necessity of disentangling semantic and collaborative signals. Building on prior work~\cite{oyama-etal-2023-norm, kurita-etal-2023-contrastive}, we hypothesize that the norm of an embedding vector serves as a proxy for the information capacity it encodes. Under this assumption, we examine the norm distributions of embedding vectors from both the tokenizer model (RQ-VAE~\cite{RQ-VAE}) and the recommender model (T5~\cite{T5}) within a representative generative recommendation framework (TIGER~\cite{TIGER}).

Specifically, for an item $i$ with semantic ID $[c_1,\ldots, c_L]$, we perform lookup operations on the codebooks $\{\bm{\mathcal{C}}_l\}_{l=1}^{L}$ and the token embedding table $\bm{\mathcal{O}}$ to obtain the code embedding sequence and token embedding sequence, each of length $L$. For each embedding sequence, we compute the norm of each embedding vector. This process is repeated for all items, and the average norm value is then calculated across the entire set. Ultimately, we obtain two vectors of length $L$, which are used to plot the histograms in Figure~\ref{fig:code_token} ($L=4$), serving as measures of information capacity across the indices of the semantic IDs.

We observe that code embeddings demonstrate a pronounced hierarchical decay, unlike the more uniform distribution of token embeddings. Given that code embeddings mainly represent semantic information whereas token embeddings integrate both semantic and collaborative information, this difference highlights the fundamentally distinct characteristics of these two signal types. This discrepancy gives rise to conflicting optimization objectives, as semantic signals tend to induce a hierarchical decay in information capacity across the indices of semantic IDs, whereas collaborative signals do not. The associated risk—that the model’s inability to effectively represent either signal ultimately impairs recommendation performance—motivates us to disentangle semantic and collaborative signals.

%% file: 4_method.tex
\label{method}
\begin{figure*}[t]
    \centering
    \includegraphics[width=\linewidth]{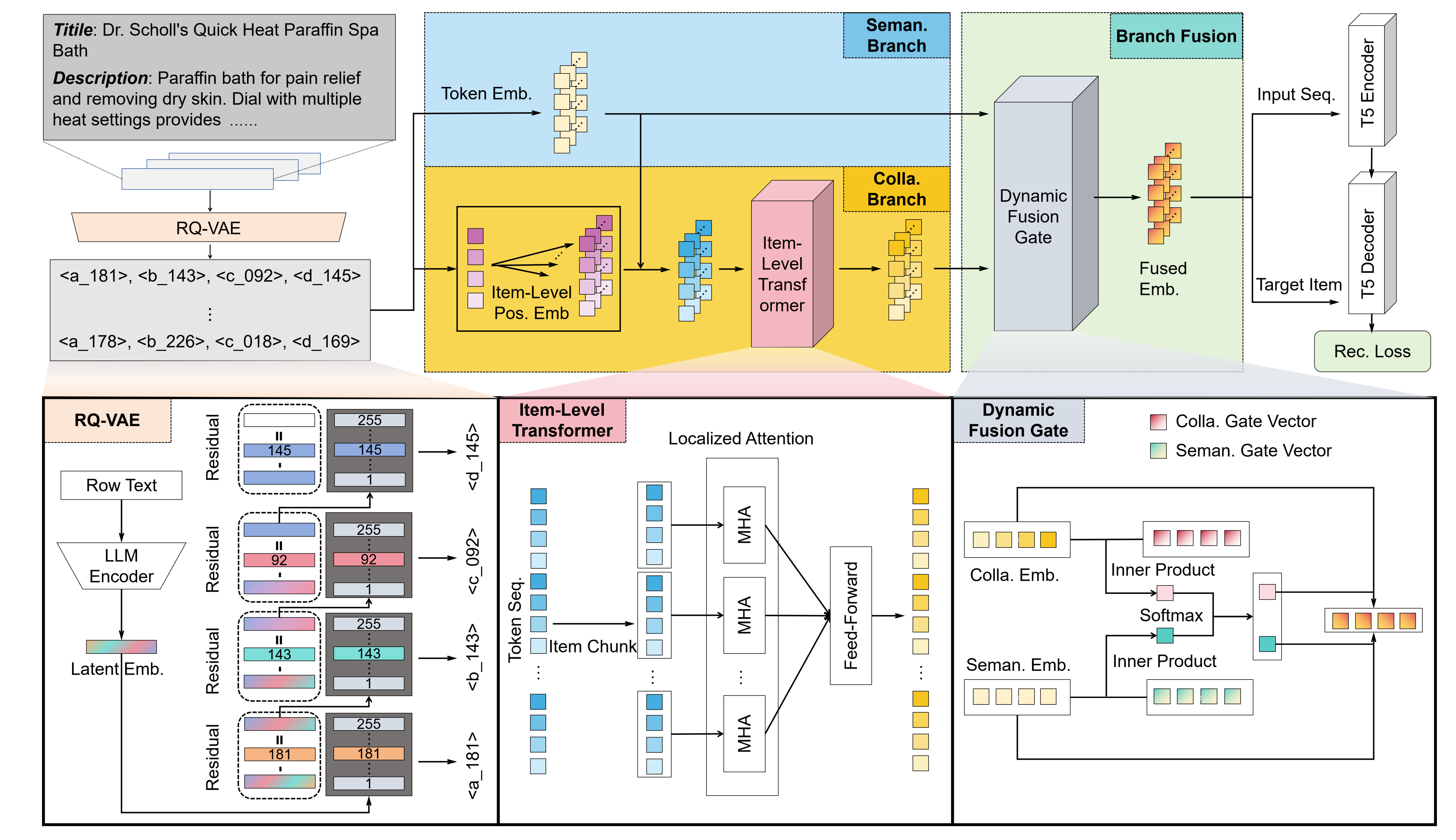}
    \caption{Overview of the proposed DiscRec framework. The upper part illustrates the generative recommendation pipeline, which sequentially employs tokenization and recommendation. Specifically, we adopt a dual-branch module for recommendation: a collaborative branch (centered on the item-level position embedding and the item-level Transformer) and a semantic branch. These two branches are adaptively fused via a dynamic fusion gate. The three subfigures below provide detailed structures of the RQ-VAE, the item-level Transformer, and the dynamic fusion gate, respectively.}
    \label{fig:method_overall}
\end{figure*}

In this section, we first provide a detailed introduction to the proposed  Disentangled Collaborative-Semantic Modeling (DiscRec) framework. 
As shown in Figure~\ref{fig:method_overall}, our DiscRec framework consists of two key modules: the \textit{Item-level Position Embedding}, which incorporates item-aware information into the input representations, and the \textit{Dual-Branch Module}, which disentangles and adaptively fuses semantic and collaborative signals. We first describe each component in detail, followed by a complexity analysis.

\subsection{Disentangled Collaborative-Semantic
Modeling}

DiscRec is designed to address two key challenges in token-level generative recommendation: item-token misalignment and semantic–collaborative signal entanglement. To this end, we introduces two complementary modules. The \textit{Item-level Position Embedding} encodes structural information about each item token's position, enabling the model to distinguish tokens across items. The \textit{Dual-Branch Module} separately models semantic and collaborative signals via parallel encoding and later fuses them adaptively at the embedding level. The following sections provide a detailed explanation of each component.

\vspace{+5pt}
\subsubsection{Item-level Position Embedding}

The key to incorporating item awareness into the architecture is enabling the model to distinguish tokens from different items rather than treating them equivalently. A straightforward solution would be to assign a unique embedding to each item; however, this approach lacks scalability, as will be discussed in the experimental section. Therefore, we adopt a lightweight alternative by constructing a dedicated \textit{Item-level Position Embedding (IPE)} table $\mathcal{V}\in\mathbb{R}^{(L+2)\times D}$. Specifically, the first $L$ rows correspond to the $L$ tokens of each item, while the last two rows respectively correspond to the special tokens [EOS] and [BOS]. For the token sequence $X$ and $Y$, the corresponding position embedding sequence is respectively defined as follows:
\begin{align}
\bm{V}^X &= [\bm{v}_1,\bm{v}_2,\ldots, \bm{v}_{L-1}, \bm{v}_{L}, \bm{v}_{L+1}], \label{eq:pos_emb_X} \\ 
\bm{V}^Y &= [\bm{v}_{L+2},\bm{v}_1, \ldots, \bm{v}_L], \label{eq:pos_emb_Y}
\end{align}
where $\bm{v}_l$ represents the $l$-th row of the embedding table $\mathcal{V}$. Notably, this embedding assignment is explicitly determined by the token’s position within the corresponding semantic ID and is independent of the item itself. By sharing these embeddings across different items, the model can efficiently capture the item-level structural information embedded within the input token sequences.

\vspace{+5pt}
\subsubsection{Dual-Branch Module}
Overall, this module aims to leverage both the original token embeddings (Equations~\eqref{eq:token_emb_X} and \eqref{eq:token_emb_Y}) and the introduced position embeddings (Equations~\eqref{eq:pos_emb_X} and \eqref{eq:pos_emb_Y}) to reconstruct the inputs to the encoder and decoder (Equations~\eqref{eq:hidden_encoder} and \eqref{eq:hidden_decoder}), thereby achieving the explicit disentanglement of the two types of signals. We denote the whole module as $\mathcal{M}$, and the resulting disentangled hidden representations of encoder and decoder are computed as follows:
\begin{align}
\bm{H}^{\text{Enc}}_{\text{Dis}}&=\text{Encoder}_{\mathcal{R}}(\mathcal{M}(\bm{E}^X,\bm{V}^X)), \label{eq:enc_hidden}\\
\bm{H}^{\text{Dec}}_{\text{Dis}} &=\text{Decoder}_{\mathcal{R}}(\bm{H}^{\text{Enc}}_{\text{Dis}},\mathcal{M}(\bm{E}^Y,\bm{V}^Y)).\label{eq:dec_hidden} 
\end{align}

Below, we present the formulation of $\mathcal{M}(\bm{E}^X,\bm{V}^X)$; the computation of $\mathcal{M}(\bm{E}^Y,\bm{V}^Y)$ can be derived in a similar manner.
The dual-branch module $\mathcal{M}$ consists of three components: a semantic branch, a collaborative branch, and an adaptive fusion mechanism. Each component is introduced in detail below:

\vspace{+5pt}
\textbf{Semantic Branch.} 
This branch is dedicated to extracting semantic signals from the semantic IDs generated during the tokenization process. Since it is independent of collaborative information, it simply outputs the original token embeddings $\bm{E}^X$ without any additional processing:
\begin{equation}
\bm{B}^{\text{Seman}} = \bm{E}^X \in\mathbb{R}^{(LT+1)\times D},
\end{equation}
where $\bm{B}^{\text{Seman}}$ represents the output of the semantic branch.

\vspace{+5pt}
\textbf{Collaborative Branch.}
This branch aims to capture collaborative signals by modeling the sequential transition patterns inherent in user interactions. To achieve this, both the original token embeddings $\bm{E}^X$ and the introduced position embeddings $\bm{V}^X$ are jointly fed into a Transformer (multi-head self-attention\footnote{We adopt bidirectional self-attention on the encoder side ($\mathcal{M}(\bm{E}^X,\bm{V}^X)$) and unidirectional attention on the decoder side ($\mathcal{M}(\bm{E}^Y,\bm{V}^Y))$.} and feed-forward network), which facilitates the integration and aggregation of collaborative information. This process can be expressed as:
\begin{equation}
\bm{B}^{\text{Colla}} = \text{Transformer}(\bm{E}^X+\bm{V}^X)\in \mathbb{R}^{(LT+1)\times D},
\end{equation}
where $\bm{B}^{\text{Colla}}$ represents the output of the collaborative branch.

However, collaborative information is at the item granularity, and the original self-attention mechanism does not align well with this requirement. Therefore, we modify the Transformer’s standard attention mechanism to a localized attention restricted to tokens within the same item, enabling more effective aggregation of collaborative signals by the \textit{Item-Level Transformer}. Specifically, this attention operation can be represented as:
\begin{equation}\label{eq:local_attn}
\text{Attn}(Q,K,V)=\text{softmax}(\frac{QK^{\top}}{\sqrt{d_k}}+W)V.
\end{equation}
Here, \(Q\), \(K\), \(V\) represent the queries, keys, values, respectively, and \(\sqrt{d_k}\) denotes the scaling factor corresponding to the dimension of the keys. $W\in\mathbb{R}^{(LT+1)\times(LT+1)}$ is a specially designed mask introduced to enforce the desired attention restriction within tokens of the same item, defined as:
\begin{equation}\label{eq:mask}
W[i][j] = 
\begin{cases}
0, & \text{if tokens } i \text{ and } j \text{ belong to the same item}, \\
-\infty, & \text{otherwise},
\end{cases}
\end{equation}
where $W[i][j]$ denotes the value at the $i$-th row and $j$-th column of the mask $W$. Notably, the special tokens [EOS] and [BOS] are considered as belonging to the nearest adjacent item for the purpose of this masking. This localized attention mechanism ensures that the model focuses on intra-item token interactions, thereby enhancing the efficacy of collaborative information aggregation.

\vspace{+5pt}
\textbf{Branch Fusion.}
To preserve the interaction between the two signal types at higher layers of the model, disentanglement is applied exclusively at the embedding level. The outputs of the two branches are then fused based on the \textit{Dynamic Fusion Gate} to produce a unified representation for downstream tasks. Specifically, two gate vectors, $\bm{G}^\text{Seman}$ and $\bm{G}^\text{Colla}$, are learned to adaptively balance and integrate the contributions from each branch. The weights are computed as the inner products between the branch outputs and their corresponding gating vectors, followed by a softmax normalization. This computation process can be formulated as:
\begin{equation}\label{eq:BG_simplified1}
\bm{B} = [\bm{B}^{\text{Seman}}, \bm{B}^{\text{Colla}}]\in\mathbb{R}^{(LT+1)\times 2\times D},
\end{equation}
\begin{equation}\label{eq:BG_simplified2}
\bm{G} = [\bm{G}^{\text{Seman}}, \bm{G}^{\text{Colla}}]\in\mathbb{R}^{1\times 2\times D},
\end{equation}
\begin{equation}\label{eq:Score_out}
\bm{S} = \text{softmax}(\bm{B} \cdot \bm{G})\in\mathbb{R}^{(LT+1)\times 2 \times 1}.
\end{equation}
\begin{equation}\label{eq:DBM_out}
\mathcal{M}(\bm{E}^X,\bm{V}^X) = \bm{S}^{\top}\bm{B}\in\mathbb{R}^{(LT+1)\times 1\times D},
\end{equation}
where $\cdot$ denotes inner product,  and $\bm{S}$ represents the fusion weights of the branch outputs.

\subsection{Training Procedure}
\begin{algorithm}[t]
\caption{Two-stage training procedure of the DiscRec}
\label{alg:training}
\begin{algorithmic}[1]
\While{not converged}\Comment{\textbf{Tokenization stage.}}
\State Compute $\mathcal{L}_\mathcal{T}$ according to Equation~\eqref{eq:loss_token};
\State Update the parameters of $\mathcal{T}$;
\EndWhile
\While{not converged}\Comment{\textbf{Recommendation stage.}}
\State Apply tokenization by $\mathcal{T}$ to get $X, Y$;
\State Compute $\mathcal{M}(\bm{E}^X,\bm{V}^X)$ according to Equation~\eqref{eq:DBM_out};
\State Compute $\mathcal{M}(\bm{E}^Y,\bm{V}^Y)$ similarly;
\State Compute $\bm{H}^{\text{Enc}}_{\text{Dis}}$ according to Equation~\eqref{eq:enc_hidden};
\State Compute $\bm{H}^{\text{Dec}}_{\text{Dis}}$ according to Equation~\eqref{eq:dec_hidden};
\State Compute $\mathcal{L}_\mathcal{R}$ according to Equation~\eqref{eq:loss_rec};
\State Update the parameters of $\mathcal{R}$;
\EndWhile
\end{algorithmic}
\end{algorithm}

Algorithm~\ref{alg:training} outlines the overall training process of the proposed DiscRec framework, which consists of two sequential stages: the tokenization stage and the recommendation stage.

In the tokenization stage (lines 1–4), the tokenizer model $\mathcal{T}$ is trained to discretize item representations into semantic IDs.
At each iteration, the tokenization loss $\mathcal{L}_\mathcal{T}$ is computed (line 2) and used to update the parameters of $\mathcal{T}$ (line 3) until convergence.

In the recommendation stage (lines 5–13), the learned tokenizer $\mathcal{T}$ is first applied to convert historical sequences and target items into tokenized inputs $X$ and $Y$ (line 6). Then, the dual-branch module $\mathcal{M}$ is used to compute disentangled representations for both the encoder and decoder sides (lines 7–8). These representations are used to construct the encoder and decoder hidden states (lines 9–10), which are then used to calculate the recommendation loss $\mathcal{L}_\mathcal{R}$ (line 11). Finally, the parameters of the recommender model $\mathcal{R}$ are updated using gradient-based optimization (line 12) until convergence.

\subsection{Complexity Analysis}

To further quantify the efficiency of the DiscRec framework, we provide a detailed analysis of its time complexity. Let $D$ denote the model dimension, $K$ the size of each codebook, $L$ the number of codebooks, and $T$ the sequence length.

\textbf{Tokenization Phase.} For a single item, the encoder and decoder layers incur a time complexity of $O(D^2)$. The search operation of each codebook introduces a cost of $O(KD)$, and the computation of the quantization loss adds an additional $O(D)$. Given $L$ codebooks, the intermediate step results in a total complexity of $O(LKD + LD)$. Furthermore, computing the overall reconstruction loss incurs an additional cost of $O(D)$. Therefore, the total complexity of the tokenization phase for a single item is $O(D^2 + LKD)$, and for a sequence of $T$ items, the overall complexity is $O(TD^2 + TLKD)$.

\textbf{Recommendation Phase.} The primary computational cost arises from the self-attention and feed-forward layers, which have a complexity of $O(T^2D + TD^2)$. In addition, the recommendation loss calculation incurs $O(TLKD)$. The collaborative branch contributes $O(T^2D + TD^2)$, while the branch fusion process incurs a cost of $O(TLKD)$.

Therefore, the overall time complexity of DiscRec is $O(TD^2 + T^2D + TLKD)$, which is in the same order as that of the TIGER baseline. Both methods also share similar inference time complexity.

%% file: 5_exp.tex
\label{exp}
In this section, we conduct a series of experiments to answer the following research questions:\\
\textbf{RQ1:} How does the proposed DiscRec framework perform compared to existing generative recommendation methods? \\
\textbf{RQ2:} What is the contribution of each component within DiscRec to the overall performance? \\
\textbf{RQ3:} Does DiscRec effectively achieve the desired item-aware alignment and flexible disentanglement? \\
\textbf{RQ4:} How well does DiscRec generalize under diverse evaluation settings? \\
\textbf{RQ5:} Can DiscRec be further improved through alternative architectural designs?

\subsection{Experimental Setting}
\label{exp_settings}
\subsubsection{Datasets}
\label{exp_datasets}

\begin{table}[t]
\centering
\caption{Statistical details of the evaluation datasets, where ``AvgLen" represents the average length of item sequences.}
\label{exp:dataset}
\begin{tabular}{cccccc}
\hline
Datasets    & \#User & \#Item & \#Interaction & Sparsity & AvgLen \\ \hline
Beauty      & 22363  & 12101  & 198502        & 99.93\%  & 8.88   \\
Instruments & 24772  & 9922   & 206153        & 99.92\%  & 8.32   \\
Toys        & 19412  & 11924  & 167597        & 99.93\%  & 8.63   \\
Arts        & 45141  & 20956  & 390832        & 99.96\%  & 8.66   \\ \hline
\end{tabular}
\end{table}

We conduct experiments on four subsets of the Amazon Review dataset\footnote{\url{https://nijianmo.github.io/amazon/index.html}}~\cite{Amazon14,Amazon18}: Beauty, Musical Instruments, Arts, and Toys. The detailed statistics of these datasets are presented in Table~\ref{exp:dataset}. Following previous work~\cite{S3-Rec}, we apply the 5-core filtering, which removes users and items with fewer than five interactions.
To standardize training, each user's interaction history is truncated or padded to a fixed length of 20, keeping the most recent interactions.
For data splitting, we adopt the widely used \textit{leave-one-out} strategy. Specifically, for each user, the most recent interaction is held out for testing, the second most recent for validation, and the remaining interactions are used for training.

\vspace{+5pt}
\subsubsection{Baselines}
\label{exp_baselines}
The baseline methods used for comparison fall into the following two categories:
\vspace{+3pt}\\
a) \textit{Traditional recommendation methods}:
\begin{itemize}[leftmargin=*]
    \item \textbf{MF}\cite{MF} decomposes the user-item interactions into the user embeddings and the item embeddings in the latent space.
    \item\textbf{LightGCN}\cite{LightGCN} captures high-order user-item interactions through a lightweight graph convolutional network.
    \item \textbf{Caser}\cite{Caser} leverages both horizontal and vertical convolutional filters to extract sequential patterns from user behavior data, capturing diverse local features.
    \item \textbf{HGN}\cite{HGN} introduces a hierarchical gating mechanism to adaptively integrate long-term and short-term user preferences derived from historical item sequences.
    \item \textbf{SASRec}\cite{SASRec} employs self-attention mechanisms to capture long-term dependencies in user interaction history.
\end{itemize}
b) \textit{Generative recommendation methods}:
\begin{itemize}[leftmargin=*]

    \item \textbf{TIGER}\cite{TIGER} adopts the generative retrieval paradigm for sequential recommendation by constructing semantic item IDs derived from text embeddings, thereby enabling semantically meaningful item representations.
    \item \textbf{LETTER}\cite{LETTER} builds upon TIGER by integrating collaborative and diversity regularization into the tokenizer learning process, enhancing the integration of semantic and collaborative signals in the tokenization stage.
    \item \textbf{EAGER}\cite{EAGER} applies a two-stream generative recommender that completely decouples heterogeneous information into separate decoding paths for parallel modeling of semantic and collaborative signals.
\end{itemize}

\vspace{+5pt}
\subsubsection{Evaluation Metrics}
\label{exp_metrics}
To ensure a fair evaluation of sequential recommendation, we adopt two widely used metrics: Recall@$K$ and NDCG@$K$ with $K=5,10$. To eliminate sampling bias, we perform full ranking over the entire item set. Additionally, during inference, we apply a constrained decoding strategy~\cite{P5-CID} that uses a prefix tree to filter out invalid semantic ID prefixes, and set the beam size to 20.

\vspace{+5pt}
\label{exp_details}
\subsubsection{Implementation Details}

For traditional methods, we follow the implementation from~\cite{S3-Rec}. For generative methods, we adopt a unified model configuration. The recommender is based on the T5 backbone, using 4 Transformer layers with a model dimension of 128, 6 attention heads (dimension 64), a hidden MLP size of 1024, ReLU activations, and a dropout rate of 0.1. The tokenizer employs RQ-VAE for discrete semantic encoding with 4 codebooks, each containing 256 embeddings of dimension 32. We set the weighting coefficient $\beta$ (in Equation~\ref{eq:loss_token}) to 0.25. Semantic inputs to RQ-VAE are derived from item titles and descriptions processed by LLaMA2-7B~\cite{Llama7b}. For collaborative embeddings in LETTER, we follow the original setup and use pretrained SASRec embeddings. For the EAGER baseline, we modify the model dimension to 128 for fair comparison, while keeping other configurations consistent with its official implementation\footnote{EAGER's original data processing does not correctly sort user interaction histories by timestamp, potentially resulting in overestimated performance. In our experiments, we address this by applying proper chronological ordering. Further details are available at \url{https://github.com/yewzz/EAGER/issues/5}}.

For our proposed DiscRec, we instantiate it with two different tokenizers by applying it to TIGER and LETTER, denoted as \textbf{DiscRec-T} and \textbf{DiscRec-L}, respectively. The collaborative branch is configured with 2 attention heads, an attention dimension of 64, an MLP hidden size of 1024 with ReLU activations, and a dropout rate of 0.1. For training the tokenizer model, we train it for 20,000 steps using a batch size of 1,024, the AdamW~\cite{AdamW} optimizer, a learning rate of 1$e$-3, and a weight decay of 1$e$-4. For training the recommender model, we use a batch size of 256 and AdamW optimizer, tuning the learning rate in {5$e$-4, 1$e$-3} with a fixed weight decay of 1$e$-2. For all baseline methods, we follow the hyper-parameter search ranges specified in their original papers.

\begin{table*}[t]
\caption{Overall performance comparison between baselines and our proposed DiscRec. Specifically, DiscRec is applied on top of TIGER and LETTER--referred to as DiscRec-T and DiscRec-L, respectively--with ``RI'' denoting the relative improvement compared to TIGER and LETTER. The best and second-best results are highlighted in bold and underlined, respectively.}
\label{exp:main}
\resizebox{\textwidth}{!}{
\begin{tabular}{ccccccccccccccc}
\hline
                              &                          & \multicolumn{5}{c}{Traditional Method}                   &  & \multicolumn{7}{c}{Generative Method}                                                                                                                                     \\ \cline{3-7} \cline{9-15} 
\multirow{-2}{*}{Dataset}     & \multirow{-2}{*}{Metric} & MF     & LightGCN & Caser  & HGN          & SASRec       &  & EAGER & TIGER  & \cellcolor[HTML]{EFEFEF}DiscRec-T       & \cellcolor[HTML]{EFEFEF}RI (\%) & LETTER       & \cellcolor[HTML]{EFEFEF}DiscRec-L          & \cellcolor[HTML]{EFEFEF}RI (\%) \\ \hline
                              & Recall@5                 & 0.0202 & 0.0228   & 0.0279 & 0.0344       & 0.0403       &  & 0.0367 & 0.0384 & \cellcolor[HTML]{EFEFEF}0.0414       & \cellcolor[HTML]{EFEFEF}7.8     & {\ul 0.0424} & \cellcolor[HTML]{EFEFEF}\textbf{0.0472} & \cellcolor[HTML]{EFEFEF}11.3   \\
                              & Recall@10                & 0.0379 & 0.0421   & 0.0456 & 0.0564       & 0.0589       &  & 0.0534 & 0.0609 & \cellcolor[HTML]{EFEFEF}{\ul 0.0656} & \cellcolor[HTML]{EFEFEF}7.7     & 0.0612       & \cellcolor[HTML]{EFEFEF}\textbf{0.0704} & \cellcolor[HTML]{EFEFEF}15.0   \\
                              & NDCG@5                   & 0.0122 & 0.0136   & 0.0172 & 0.0214       & {\ul 0.0271} &  & 0.0266 & 0.0256 & \cellcolor[HTML]{EFEFEF}0.0270       & \cellcolor[HTML]{EFEFEF}5.5     & 0.0264       & \cellcolor[HTML]{EFEFEF}\textbf{0.0308} & \cellcolor[HTML]{EFEFEF}16.7   \\
\multirow{-4}{*}{Beauty}      & NDCG@10                  & 0.0178 & 0.0198   & 0.0229 & 0.0284       & 0.0331       &  & 0.0322 & 0.0329 & \cellcolor[HTML]{EFEFEF}{\ul 0.0348} & \cellcolor[HTML]{EFEFEF}5.8     & 0.0334       & \cellcolor[HTML]{EFEFEF}\textbf{0.0382} & \cellcolor[HTML]{EFEFEF}14.4   \\ \hline
                              & Recall@5                 & 0.0738 & 0.0757   & 0.0770 & 0.0854       & 0.0857       &  & 0.0599 & 0.0865 & \cellcolor[HTML]{EFEFEF}{\ul 0.0901} & \cellcolor[HTML]{EFEFEF}4.2     & 0.0872       & \cellcolor[HTML]{EFEFEF}\textbf{0.0932} & \cellcolor[HTML]{EFEFEF}6.9    \\
                              & Recall@10                & 0.0967 & 0.1010   & 0.0995 & 0.1086       & 0.1083       &  & 0.0745 & 0.1062 & \cellcolor[HTML]{EFEFEF}{\ul 0.1101} & \cellcolor[HTML]{EFEFEF}3.7     & 0.1082       & \cellcolor[HTML]{EFEFEF}\textbf{0.1153} & \cellcolor[HTML]{EFEFEF}6.6    \\
                              & NDCG@5                   & 0.0473 & 0.0472   & 0.0639 & 0.0723       & 0.0715       &  & 0.0506 & 0.0736 & \cellcolor[HTML]{EFEFEF}{\ul 0.0752} & \cellcolor[HTML]{EFEFEF}2.2     & 0.0747       & \cellcolor[HTML]{EFEFEF}\textbf{0.0791} & \cellcolor[HTML]{EFEFEF}5.9    \\
\multirow{-4}{*}{Instruments} & NDCG@10                  & 0.0547 & 0.0554   & 0.0711 & 0.0798       & 0.0788       &  & 0.0553 & 0.0799 & \cellcolor[HTML]{EFEFEF}{\ul 0.0817} & \cellcolor[HTML]{EFEFEF}2.3     & 0.0814       & \cellcolor[HTML]{EFEFEF}\textbf{0.0862} & \cellcolor[HTML]{EFEFEF}5.9    \\ \hline
                              & Recall@5                 & 0.0218 & 0.0253   & 0.0243 & 0.0328       & 0.0357       &  & 0.0301 & 0.0316 & \cellcolor[HTML]{EFEFEF}{\ul 0.0367} & \cellcolor[HTML]{EFEFEF}16.1    & 0.0334       & \cellcolor[HTML]{EFEFEF}\textbf{0.0403} & \cellcolor[HTML]{EFEFEF}20.7   \\
                              & Recall@10                & 0.0350 & 0.0413   & 0.0404 & 0.0417       & {\ul 0.0602} &  & 0.0451 & 0.0511 & \cellcolor[HTML]{EFEFEF}0.0583       & \cellcolor[HTML]{EFEFEF}14.1    & 0.0542       & \cellcolor[HTML]{EFEFEF}\textbf{0.0634} & \cellcolor[HTML]{EFEFEF}17.0   \\
                              & NDCG@5                   & 0.0123 & 0.0149   & 0.0150 & {\ul 0.0243} & 0.0228       & & 0.0215 & 0.0204 & \cellcolor[HTML]{EFEFEF}0.0235       & \cellcolor[HTML]{EFEFEF}15.2    & 0.0225       & \cellcolor[HTML]{EFEFEF}\textbf{0.0254} & \cellcolor[HTML]{EFEFEF}12.9   \\
\multirow{-4}{*}{Toys}        & NDCG@10                  & 0.0165 & 0.0200   & 0.0201 & 0.0272       & {\ul 0.0306} &  & 0.0268 & 0.0267 & \cellcolor[HTML]{EFEFEF}0.0304       & \cellcolor[HTML]{EFEFEF}13.9    & 0.0292       & \cellcolor[HTML]{EFEFEF}\textbf{0.0328} & \cellcolor[HTML]{EFEFEF}12.3   \\ \hline
                              & Recall@5                 & 0.0473 & 0.0464   & 0.0571 & 0.0667       & 0.0758       &  & 0.0555 & 0.0807 & \cellcolor[HTML]{EFEFEF}0.0822       & \cellcolor[HTML]{EFEFEF}1.9     & {\ul 0.0841} & \cellcolor[HTML]{EFEFEF}\textbf{0.0850} & \cellcolor[HTML]{EFEFEF}1.1    \\
                              & Recall@10                & 0.0753 & 0.0755   & 0.0781 & 0.0910       & 0.0945       &  & 0.0694 & 0.1017 & \cellcolor[HTML]{EFEFEF}0.1061       & \cellcolor[HTML]{EFEFEF}4.3     & {\ul 0.1081} & \cellcolor[HTML]{EFEFEF}\textbf{0.1110} & \cellcolor[HTML]{EFEFEF}2.7    \\
                              & NDCG@5                   & 0.0271 & 0.0244   & 0.0407 & 0.0516       & 0.0632       &  & 0.0472 & 0.0640 & \cellcolor[HTML]{EFEFEF}0.0659       & \cellcolor[HTML]{EFEFEF}3.0     & {\ul 0.0675} & \cellcolor[HTML]{EFEFEF}\textbf{0.0684} & \cellcolor[HTML]{EFEFEF}1.3    \\
\multirow{-4}{*}{Arts}        & NDCG@10                  & 0.0361 & 0.0338   & 0.0474 & 0.0595       & 0.0693       &  & 0.0518 & 0.0707 & \cellcolor[HTML]{EFEFEF}0.0736       & \cellcolor[HTML]{EFEFEF}4.1     & {\ul 0.0752} & \cellcolor[HTML]{EFEFEF}\textbf{0.0768} & \cellcolor[HTML]{EFEFEF}2.1    \\ \hline
\end{tabular}
}
\end{table*}

\subsection{Performance Comparison (RQ1)}
\label{exp_overall}

We begin by assessing the overall recommendation performance of the compared methods on all four datasets. The summarized results are presented in Table~\ref{exp:main}, yielding the following key observations:
\begin{itemize}[leftmargin=*]
\item 
Applying DiscRec to both TIGER and LETTER consistently improves recommendation performance (DiscRec-T$>$TIGER, DiscRec-L$>$LETTER), with DiscRec-L achieving the best overall results. These gains stem from the incorporation of item-awareness within the framework, which aligns the Transformer-based recommender architecture more closely with the recommendation task, and from explicit signal disentanglement that facilitates more effective and coordinated learning of semantic and collaborative information.

\vspace{+5pt}
\item 
Traditional recommendation methods generally underperform compared to generative approaches, primarily due to their reliance on ID embeddings, which limits their ability to capture hierarchical semantic structures at multiple levels of granularity. In contrast, generative models enhanced with RQ-VAE are capable of modeling such structures, allowing for more fine-grained distinctions between semantically similar items through richer, continuous representations.

\vspace{+5pt}
\item 
LETTER outperforms TIGER, largely due to the additional supervision introduced by incorporating collaborative signals during the tokenization stage. Specifically, LETTER constructs an auxiliary loss based on collaborative embeddings, effectively bridging the gap between items with similar semantics but divergent interaction patterns.

\vspace{+5pt}
\item
Although EAGER adopts a two-stream generative design for disentanglement, its performance is limited; on the Instruments dataset, it is even outperformed by traditional baselines like MF. This suggests that processing collaborative and semantic signals independently may lead to over-separation, hindering their integration and diminishing the model’s ability to capture their complementary strengths. 

\end{itemize}

\subsection{Ablation Study (RQ2)}
\label{exp_ablation}

To validate the design rationale behind DiscRec, we perform a thorough evaluation by systematically ablation of each key component, yielding a set of model variants. Specifically, starting from DiscRec-T, we introduce the following variants for comparison:
\begin{itemize}[leftmargin=*]
\item \textbf{w/o IPE}: This variant removes the item-level position embedding from the collaborative branch, relying solely on the token embeddings to model collaborative signals.

\vspace{+2pt}
\item \textbf{w/o TF}: This variant eliminates the whole introduced Transformer equipped with localized attention operation.

\vspace{+2pt}
\item \textbf{w/o Gating}: This variant eliminates the adaptive gating fusion mechanism, replacing it with a direct summation of the two branch representations.
\end{itemize}

In addition to the above basic variants, we further design several alternative variants based on TIGER to explore potential strategies for incorporating item-awareness:
\begin{itemize}[leftmargin=*]
\item \textbf{w/ PE}: This variant adopts standard learnable positional embeddings, implemented as an embedding table with a size equal to the maximum length of the input token sequence.

\vspace{+2pt}
\item \textbf{w/ IE}: This variant introduces an additional item ID embedding, which is concatenated to the end of each item's original token embedding. The size of the embedding table corresponds to the total number of unique items.

\vspace{+2pt}
\item \textbf{w/ IPE}: This variant directly applies the item-level positional embedding in DiscRec by summation without any further modifications.
\end{itemize}

Lastly, to rule out the impact of increased model capacity, we introduce an additional variant:
\begin{itemize}[leftmargin=*]
\item \textbf{w/ 5-Layer}: The number of transformer layers in TIGER is increased to 5, resulting in a parameter increase greater than that introduced by DiscRec to ensure a fair comparison.
\end{itemize}

\vspace{+5pt}
Table~\ref{exp:abl} illustrates the comparison results on Beauty, from which we draw the following observations:

\begin{itemize}[leftmargin=*]

\item The removal of item-level positional embedding, Transformer with localized attention and adaptive gating all leads to noticeable performance degradation, confirming the effectiveness of these specific design choices. Among them, w/o IPE and w/o Attn results in a more substantial decline in performance. Since both components are applied within the collaborative branch, this finding underscores the critical importance of effectively leveraging collaborative signals.

\vspace{+5pt}
\item Among the three TIGER variants incorporating item-awareness, both w/ IPE and w/ IE lead to consistent performance improvements, whereas w/ PE results in negligible change. This suggests that the former two approaches effectively introduce item-level structural patterns, thereby improving alignment with the recommendation task. Although w/ IE yields greater performance gains, its reliance on a large item embedding table imposes scalability constraints. In contrast, the proposed IPE offers a more scalable and cost-effective solution, thereby enhancing practicality in large-scale recommendation scenarios.

\vspace{+5pt}
\item Simply increasing the recommender model capacity in TIGER does not lead to significant performance gains, likely due to its misaligned architecture and the entanglement of information. This underscores that the improvements achieved by our method stem not merely from increased parameters, but from enhanced task alignment and effective disentanglement of semantic and collaborative signals.
\end{itemize}

\begin{table}[t]
\centering
\caption{Ablation study of DiscRec-T on Beauty.}
\label{exp:abl}
\begin{tabular}{ccccc}
\hline
Method            & Recall@5        & NDCG@5          & Recall@10       & NDCG@10         \\ \hline
DiscRec-T        & \textbf{0.0414} & \textbf{0.0270}  & \textbf{0.0656} & \textbf{0.0348} \\
\textit{w/o IPE}  & 0.0382          & 0.0246         & 0.0619          & 0.0322        \\
\textit{w/o TF}   & 0.0377         & 0.0253          & 0.0605       & 0.0327          \\ 
\textit{w/o Gating} & 0.0403          & 0.0262          & 0.0629          & 0.0334          \\ \hline
TIGER             & {0.0381}    & 0.0243          & 0.0593          & 0.0311          \\
\textit{w/ PE}    & 0.0369          & 0.0237          & 0.0597          & 0.0310           \\
\textit{w/ IE}    & 0.0396 & 0.0257 & 0.0619 & 0.0329 \\
\textit{w/ IPE}   & 0.0372          & {0.0249}    & {0.0600}      & {0.0322}    \\ 
\textit{w/ 5-Layer}    & 0.0367          & 0.0240           & 0.0590           & 0.0312          \\ \hline
\end{tabular}
\end{table}

\subsection{Effectiveness Analysis (RQ3)}
\label{exp_modelvalidate}

\begin{figure}[t]
  \centering
  \includegraphics[width=\linewidth]{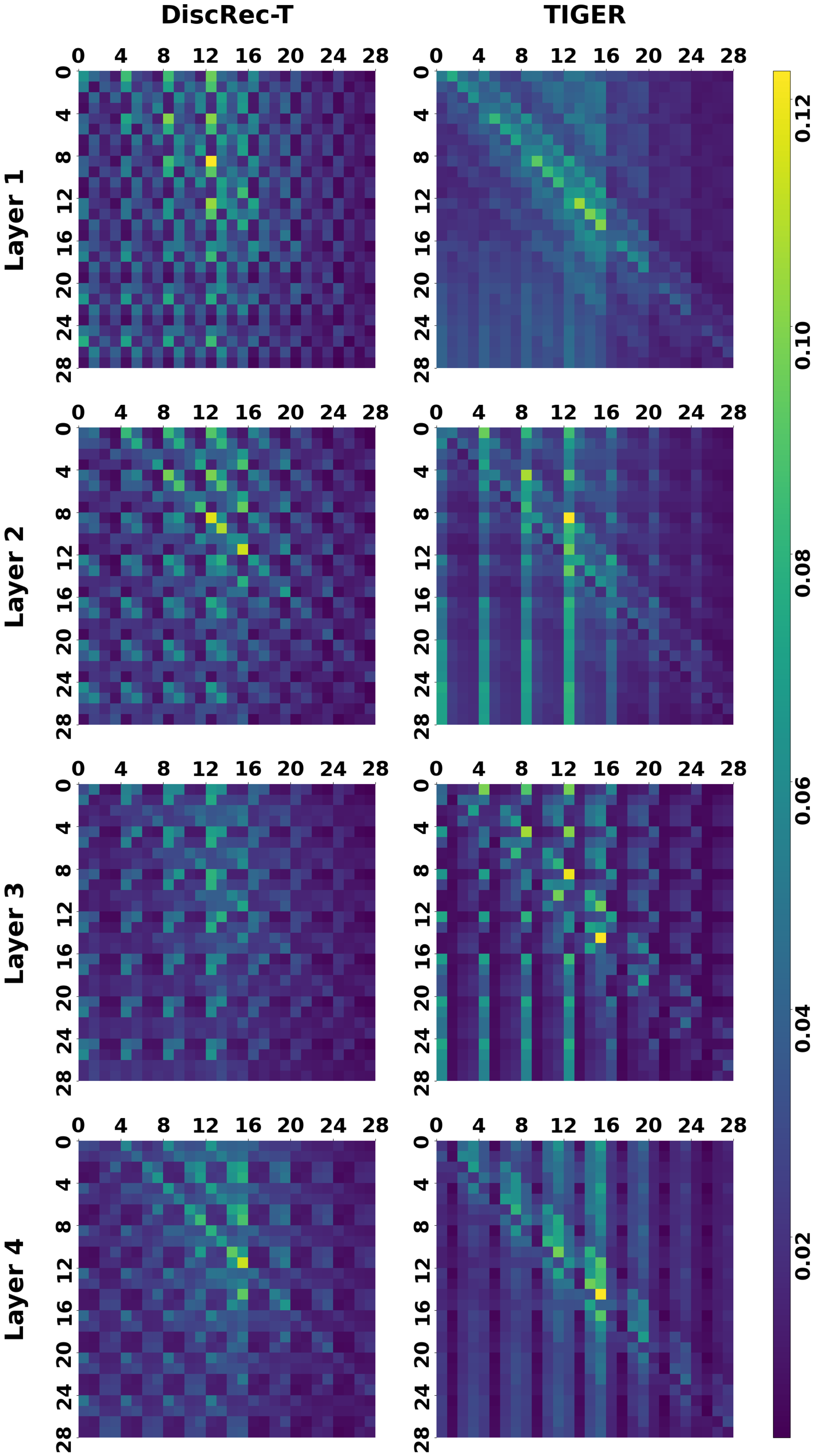}
  \caption{Layer-wise attention heatmaps for the four Transformer layers of DiscRec-T and TIGER, cropped to 28×28 to minimize the impact of padding.}
  \label{fig:heatmap1}
\end{figure}

\begin{figure}[t]
  \centering
  \includegraphics[width=\linewidth]{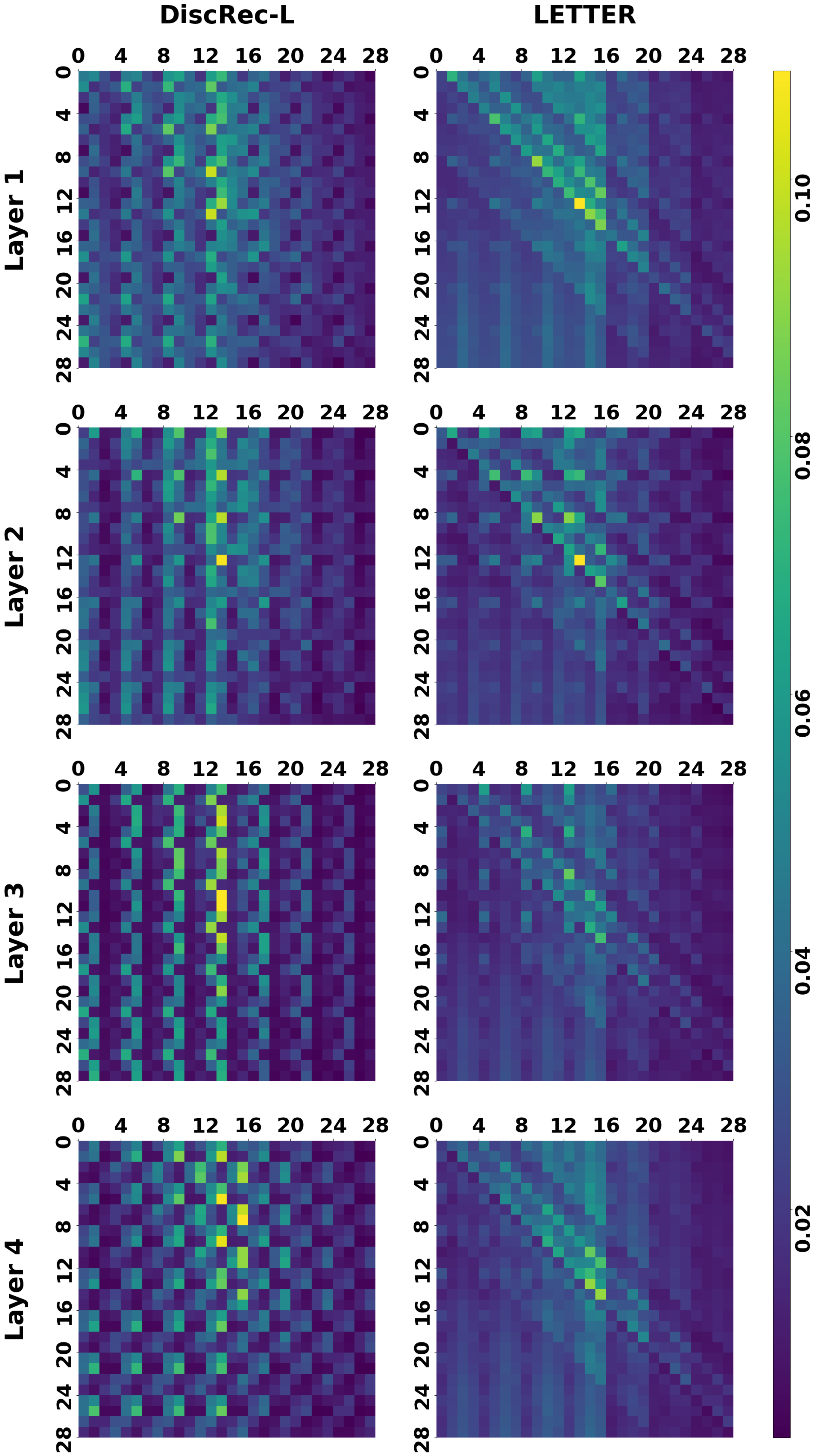}
  \caption{Layer-wise attention heatmaps for the four Transformer layers of DiscRec-L and LETTER, cropped to 28×28 to minimize the impact of padding.}
  \label{fig:heatmap2}
\end{figure}

\begin{figure}[t]
    \centering
    \includegraphics[width=\linewidth]{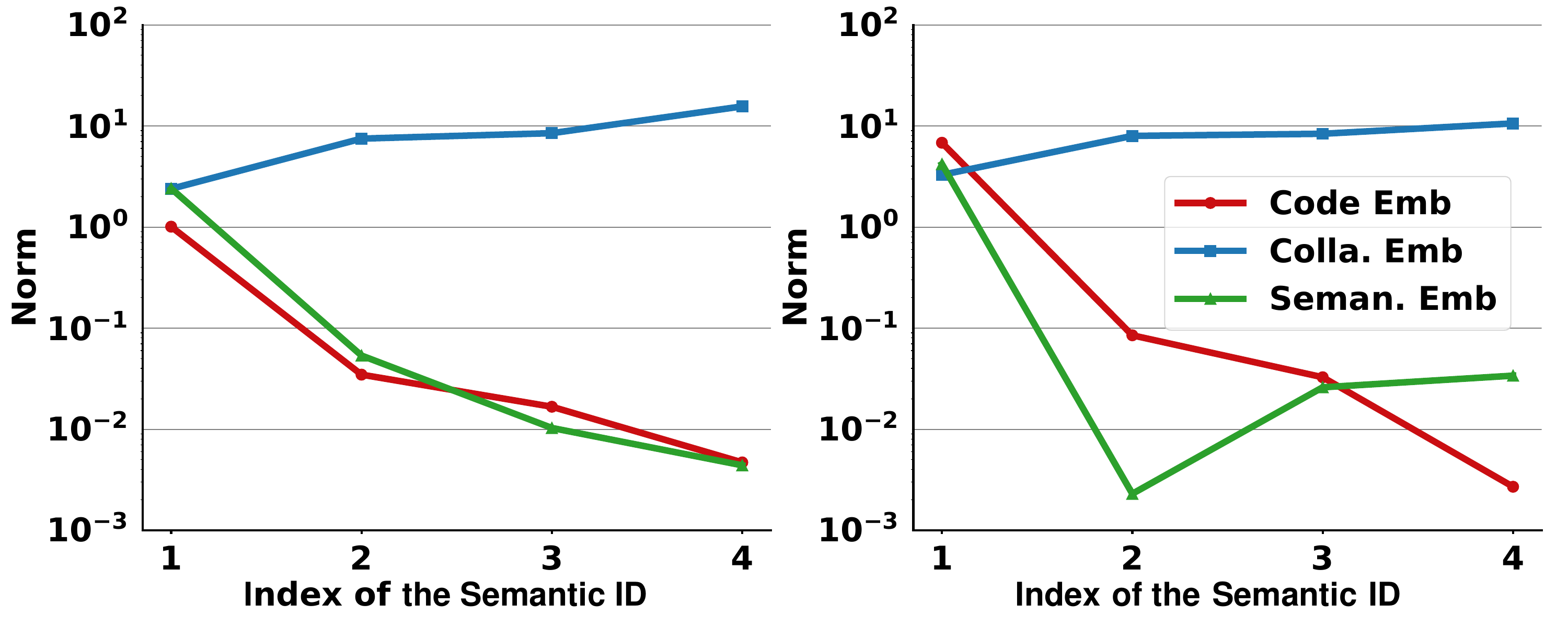}
    \caption{Embedding norm distribution comparison of code embeddings from the tokenizer, semantic token embeddings, and collaborative token embeddings from the recommender.}
    \label{fig:exp_norm}
\end{figure}

We propose DiscRec to address two fundamental challenges: token-item misalignment, and semantic–collaborative signal entanglement. Therefore, we next evaluate the extent to which these challenges have been successfully mitigated.

\vspace{+5pt}
\subsubsection{Item-Aware Alignment}

First, to validate the effectiveness of DiscRec in achieving item-aware alignment, we visualize the attention distributions across the four encoder layers of the recommender model. Using the Beauty test set, we average attention scores over six heads per layer and present the results as heatmaps. Since the average user interaction length in the test set of Beauty is 7.19--shorter than the maximum sequence length of 20--many padding tokens produce extensive low-activation regions in the heatmaps, complicating interpretation. To enhance clarity, we crop all heatmaps to a fixed size of 28×28 based on the average interaction length. For comparison, we include the TIGER and LETTER baseline, which excludes all DiscRec-specific components. 

As shown in Figure~\ref{fig:heatmap1} and Figure~\ref{fig:heatmap2}, the attention heatmaps across all four encoder layers of DiscRec-T and DiscRec-L display a clear 4×4 grid-like segmentation pattern. This regularity results from discretizing each item into four tokens, which induces structured attention distributions aligned with item-level granularity. In contrast, such patterns are not evident in the corresponding layers of the TIGER and LETTER baselines. These observations suggest that DiscRec effectively models fine-grained item-level structure throughout the encoding process, thereby improving task alignment and enhancing the extraction of collaborative signals.

\vspace{+5pt}
\subsubsection{Semantic-Collaborative Signal Disentanglement with Flexible Fusion}
To assess the effectiveness of DiscRec in achieving signal disentanglement and flexible fusion, we next examine the distribution of semantic and collaborative signals. Specifically, we focus on the distribution of three types of embeddings in relation to the semantic ID index: code embeddings produced by the tokenizer model (Code Emb.), semantic token embeddings from the semantic branch (Seman. Emb.), and collaborative token embeddings from the collaborative branch (Colla. Emb.). Using the Beauty dataset, we compute the average embedding norm as a measure of information capacity and perform this analysis for both DiscRec-T and DiscRec-L.

As illustrated in Figure \ref{fig:exp_norm}, two key observations emerge:
\begin{itemize}[leftmargin=*]
\item 
For both DiscRec-T and DiscRec-L, the semantic token embeddings exhibit distribution patterns similar to those of the code embeddings, showing a consistent decline across token indices—a characteristic inherent to residual quantization. This indicates that the semantic information originally encoded in the semantic ID sequences has been successfully disentangled and captured by the semantic branch.

\vspace{+5pt}
\item 
In contrast, the collaborative token embeddings exhibit a markedly different distribution pattern, increasing with indices of the semantic ID. This suggests that \textbf{during item decoding, the model gradually shifts its focus from semantic signals toward collaborative signals}. This observation aligns with our hypothesis in Figure~\ref{fig:code_token} that \textbf{semantic and collaborative signals display significantly distinct distribution patterns in generative recommendation}. Moreover, these findings confirm that DiscRec could effectively disentangle two types of signals.
\end{itemize}

\subsection{Generalization Analysis (RQ4)}
\label{exp_generalization}

\begin{figure}[t]
\centering
{\includegraphics[width=\linewidth]{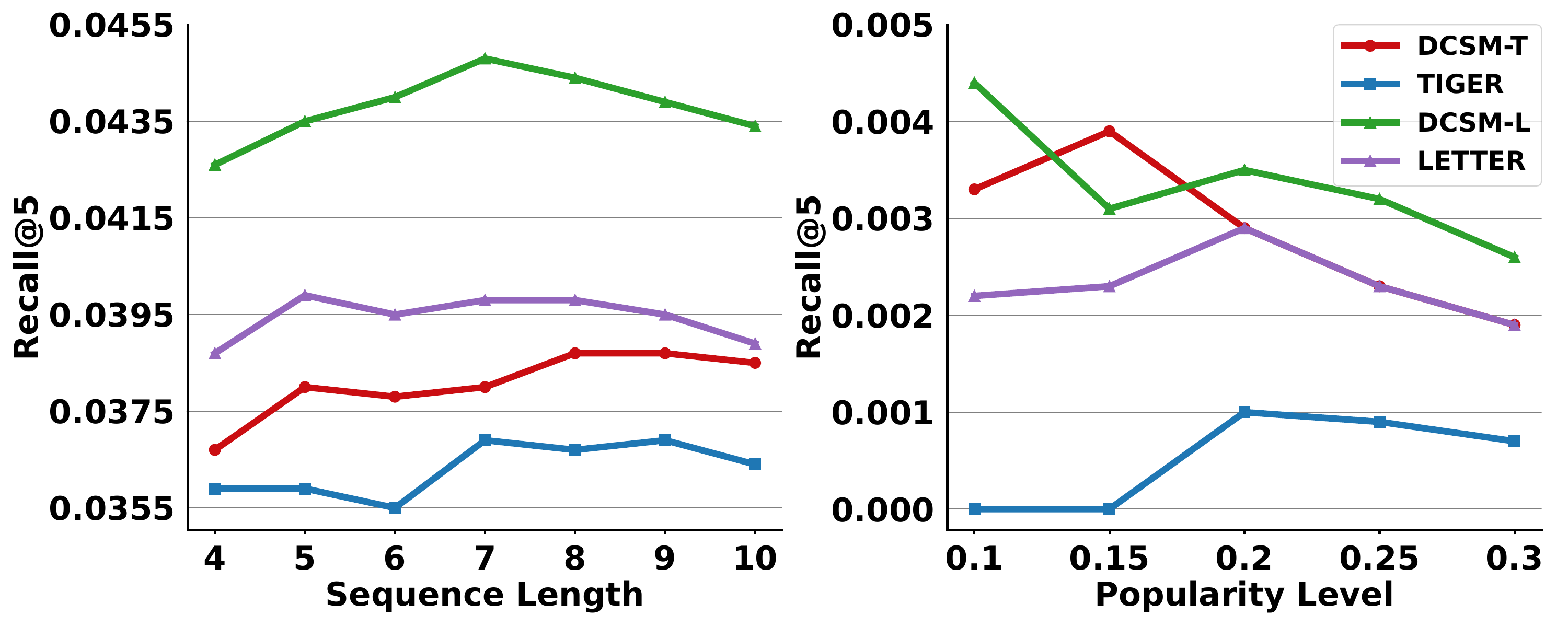}}
\caption{Performance comparison on different sequence lengths and popularity levels on Beauty.}
\label{fig:gen}
\end{figure}

\begin{table}[t]
\centering
\caption{Performance comparison under user-level random split on Beauty.}
\label{tab:user_split}
\begin{tabular}{ccccc}
\hline
Method      & Recall@5        & NDCG@5          & Recall@10       & NDCG@10         \\ \hline
DiscRec-T  & 0.0786 & 0.0530  & {0.1157} & {0.0649} \\
TIGER       & 0.0754          & 0.0509          & 0.1112          & 0.0625          \\ \hline
DiscRec-L & {0.0847} & {0.0581} & {0.1208} & {0.0697} \\
LETTER      & 0.0796          & 0.0544          & 0.1165          & 0.0662          \\ \hline
\end{tabular}
\end{table}

To evaluate DiscRec’s generalization, we test it on the Beauty dataset across three scenarios: varying input sequence lengths, different item popularity levels, and alternative data splits:

\begin{itemize}[leftmargin=*]
\item \textbf{Sequence Length}: To evaluate robustness across limited user history lengths, we group test samples by input sequence lengths from 4 to 10, noting that the minimum length of 4 results from our 5-core preprocessing rules, and assess model performance within each group.

\vspace{+2pt}
\item \textbf{Popularity Level}: To assess robustness across varying item popularity levels, we group test samples according to the popularity of their target items. We then evaluate on subsets where the target items fall within the lowest 10\%, 15\%, 20\%, 25\%, and 30\% of the popularity distribution.

\vspace{+2pt}
\item \textbf{Data Split}: To evaluate robustness under different data partitioning schemes, we replace the original leave-one-out split with a user-level random split. Specifically, the dataset is shuffled at the user level and divided into training, validation, and test sets in an 8:1:1 ratio. All models are then retrained from scratch using this new split.
\end{itemize}

The results are shown in Figure~\ref{fig:gen} and Table~\ref{tab:user_split}, from which we make the following observations:
\begin{itemize}[leftmargin=*]

\item 
Across varying sequence lengths and item popularity levels, DiscRec consistently achieves performance improvements over both TIGER and LETTER. This can be attributed to its more balanced integration of collaborative and semantic signals, enabling robust gains across diverse scenarios.

\vspace{+5pt}
\item 
Under the user-based split, DiscRec continues to deliver consistent performance improvements, suggesting that its effectiveness is not contingent on specific user behavior patterns. This highlights the model’s robustness and strong generalization capability across diverse user distributions.
\end{itemize}

\subsection{Potential Design Analysis (RQ5)}
\label{exp_indepth}

To further investigate the potential extensions of DiscRec’s architectural components, we conduct exploratory experiments with four enhanced variants of the framework:
\begin{itemize}[leftmargin=*]
\item \textbf{AllLayer}: This variant extends the application of DiscRec beyond the embedding layer to all Transformer layers.

\vspace{+2pt}
\item \textbf{OneQuery}: Inspired by Q-former designs, this variant replaces the localized attention in the collaborative branch of DiscRec with an attention mechanism that uses a single learnable query embedding for all tokens.

\vspace{+2pt}
\item \textbf{TokenAvg}: This variant introduces an additional step to the Transformer output in the collaborative branch, where token representations corresponding to the same item are averaged to form a unified item-level representation.

\vspace{+2pt}
\item \textbf{SelfGating}: This variant modifies DiscRec’s adaptive gating mechanism by adopting a self-gating strategy inspired by HGN~\cite{HGN}. Specifically, it computes gating vectors via an MLP with sigmoid activation applied to the outputs of the two branches, and uses element-wise multiplication to modulate the original representations.
\end{itemize}

The performance of these four variants is presented in Table~\ref{exp:explore}, and we make the following observations:
\begin{itemize}[leftmargin=*]
\item Applying DiscRec to all Transformer layers (AllLayer) results in a significant performance decline, falling below the TIGER baseline, which suggests that decoupling is most suitable at the embedding layer. This is likely because the higher layers primarily facilitate information interaction, where decoupling may be detrimental.

\vspace{+5pt}
\item OneQuery performs slightly worse than TIGER, suggesting that learning a unified collaborative query embedding from sequential patterns is challenging. This highlights the advantage of directly using self-attention for modeling collaborative signals. TokenAvg also shows slightly inferior performance compared to DiscRec-T, indicating that preserving token-level personalized collaborative information is more effective than simple averaging-based aggregation.

\vspace{+5pt}
\item 
Self-Gating shows no significant performance difference compared to DiscRec-T, suggesting that directly learning a gating vector is a simple yet effective approach for adaptively fusing the two types of information.
\end{itemize}

\begin{table}[t]
\centering
\caption{Exploratory experimental results on Beauty.}
\label{exp:explore}
\begin{tabular}{ccccc}
\hline
Method     & Recall@5        & NDCG@5         & Recall@10       & NDCG@10         \\ \hline
DiscRec-T & \textbf{0.0414} & \textbf{0.0270} & \textbf{0.0656} & \textbf{0.0348} \\
TIGER             & {0.0381}    & 0.0243          & 0.0593          & 0.0311          \\ \hline
AllLayer      & 0.0313         & 0.0191        & 0.0504         & 0.0252         \\
OneQuery      & 0.0351          & 0.0225         & 0.0563          & 0.0293          \\
TokenAvg      & 0.0397          & 0.0255         & 0.0614          & 0.0324          \\
SelfGating   & 0.0406          & 0.0268         & 0.0640           & 0.0343          \\ \hline
\end{tabular}
\end{table}

%% file: 6_rel.tex
\label{rel}
In this section, we navigate through existing research on sequential recommendation, generative recommendation and collaborative information modeling.

\subsection{Sequential Recommendation}

Sequential recommendation aims to predict the next item a user will interact with based on their historical behavior. Early approaches primarily utilized Markov Chains to capture item transition dynamics~\cite{FuseSim,FPMC}. Later developments introduced a range of deep learning architectures. Specifically, GRU4Rec~\cite{GRU4Rec} pioneered the use of recurrent neural networks (RNNs) in session-based recommendation, whereas Caser~\cite{Caser} adopted convolutional neural networks (CNNs) to model user behavior. More recently, methods like SASRec~\cite{SASRec} and BERT4Rec~\cite{BERT4Rec} have employed self-attention mechanisms, achieving state-of-the-art results in sequential recommendation tasks. To enrich item representations with contextual signals, FDSA~\cite{FDSA} incorporates a self-attention module that effectively captures item attribute information. In a similar vein, $\text{S}^3$-Rec~\cite{S3-Rec} enhances performance by maximizing mutual information across multiple types of contextual data. Although these methods have significantly advanced the field, they remain fundamentally discriminative in nature and thus face inherent limitations in generalization ability.

\subsection{Generative Recommendation}

Motivated by the remarkable success of generative AI, there has been a growing interest in transitioning recommendation paradigms from discriminative to generative frameworks across both academia and industry. This emerging generative recommendation paradigm typically consists of two core stages: a tokenization stage, which encodes item semantics into discrete representations, and a generation stage, which produces item predictions based on these representations. TIGER~\cite{TIGER} is among the pioneering efforts in this direction, leveraging RQ-VAE to discretize item representations into semantic tokens and employing an encoder–decoder architecture to model user behavior sequences for item generation.
Subsequent research has aimed to enhance these two stages. For the tokenization stage, several works have explored the additional integration of auxiliary information to improve the adaptability of semantic IDs~\cite{LETTER,ActionPiece,P5,P5-CID,IDGenRec,TokenRec}. For example, LETTER~\cite{LETTER} incorporates collaborative and diversity regularization into the token learning process, while ActionPiece~\cite{ActionPiece} introduces context-awareness, enabling tokens to reflect different semantics depending on their surrounding context. On the recommendation side, researchers have proposed diverse architectural designs~\cite{EAGER,LC-Rec,HSTU,SEATER} to better capture user–item interactions. For instance, EAGER~\cite{EAGER} presents a two-stream architecture that parallelly models semantic and collaborative signals. LC-Rec~\cite{LC-Rec}, on the other hand, explores the integration of large language models (LLMs) to more effectively leverage semantic IDs.

\subsection{Collaborative Information Modeling}

In traditional ID-based recommendation paradigms, the modeling of collaborative signals derived from co-occurrence patterns in user-item interactions serves as a cornerstone. Recently, with the rapid adoption of large language models (LLMs), numerous studies have explored leveraging the world knowledge and reasoning capabilities of LLMs to perform recommendation tasks. For example, TALLRec~\cite{TALLRec} fine-tunes LLMs on recommendation data to perform click-through rate (CTR) prediction tasks. BIGRec~\cite{BIGRec} first grounds LLMs in the recommendation space by fine-tuning them to generate meaningful tokens corresponding to items, and then retrieves actual items that align with the generated tokens. However, these methods typically represent users and items as text tokens, relying primarily on textual semantics, which inherently fail to capture collaborative signals. As a result, a growing body of research has emerged to incorporate collaborative information into LLM-based recommendation systems. For instance, CoLLM~\cite{CoLLM} integrates collaborative signals by leveraging an external traditional recommendation model and mapping the resulting representations into the input embedding space of the LLM, thereby forming collaborative-aware embeddings for downstream tasks. Similarly, LLaRA~\cite{LLaRA} incorporates ID embeddings and historical interaction sequences directly into the prompt to enhance the model's capacity for sequential recommendation.

In the aforementioned two-stage generative recommendation paradigm, efforts have also been made to leverage both collaborative and semantic signals. These approaches can be broadly categorized into two groups. The first group enhances the tokenizer with collaborative supervision signals. For instance, LETTER~\cite{LETTER} incorporates embeddings from a pre-trained ID-based collaborative filtering model as additional supervision, while ETEGRec~\cite{ETEGRec} utilizes outputs from the downstream recommender model for the same purpose. The second group adopts dual-branch architectures, where semantic and collaborative signals are modeled independently. For example, EAGER~\cite{EAGER} employs separate decoders for the modeling of semantic and collaborative tasks. SC-Rec~\cite{SC-Rec} leverages a unified LLM as a reasoning engine, performing independent task learning on both collaborative and semantic token sequences.
However, these methods fail to address the challenges of token-item misalignment and semantic–collaborative signal entanglement. that we aim to resolve in this work. In contrast, our work addresses this gap through a more in-depth exploration of this disentanglement problem in generative recommendation.

%% file: 7_con.tex
\label{con}

This paper proposed DiscRec, a novel framework for generative recommendation that addresses two fundamental challenges: token-item misalignment, and semantic–collaborative signal entanglement. Traditional token-level architectures fail to capture item-level granularity and entangle heterogeneous signals in a shared embedding space, which hinders effective learning. To tackle these issues, DiscRec introduces item-level position embeddings to capture item-aware structural information within token sequences. In addition, it employs a dual-branch architecture to disentangle semantic and collaborative signals at the embedding layer, with a gating mechanism that adaptively fuses the two representations, enabling flexible disentanglement while retaining cross-signal interactions at deeper layer. Overall, DiscRec offers a lightweight solution for achieving item-aware alignment, and flexible signal disentanglement and fusion.

In future work, we will explore several promising directions to further enhance the DiscRec framework. First, we plan to adapt the DiscRec framework to LLM-based recommendation scenarios. Additionally, we plan to investigate more advanced disentanglement strategies, such as contrastive learning or mutual information minimization, to more effectively separate semantic and collaborative signals.

%% file: 8_ai.tex
\label{ai}
We leveraged ChatGPT (OpenAI, GPT‑4‑turbo, 2025) exclusively to polish the prose of the manuscript drafted by the authors, focusing on grammar correction, clarity enhancement, and overall readability improvement. During development, it was also employed as a debugging aid for code. At no stage did ChatGPT contribute to the generation of substantive research content—this includes formulation of concepts or algorithms, implementation of code, data analysis, presentation of experimental findings, or creation of tables and figures.